\documentclass[reprint,amsfonts, amssymb, amsmath,  showkeys, nofootinbib,pra, superscriptaddress, twocolumn,longbibliography]{revtex4-2}
\usepackage{float}
\makeatletter
\let\newfloat\newfloat@ltx
\makeatother
\usepackage[english]{babel}
\usepackage[utf8]{inputenc}
\usepackage{graphics}
\usepackage{selinput}

\usepackage{braket}
\usepackage{amsthm}
\usepackage{mathtools}
\usepackage{physics}
\usepackage[dvipsnames]{xcolor}
\usepackage{graphicx}
\usepackage[left=16mm,right=16mm,top=35mm,columnsep=15pt]{geometry} 
\usepackage{adjustbox}
\usepackage{placeins}
\usepackage[T1]{fontenc}
\usepackage{lipsum}
\usepackage{csquotes}
\usepackage{mathbbol} 
\usepackage[linesnumbered,ruled,vlined]{algorithm2e}
\SetKwInput{kwInit}{Init}

\usepackage[makeroom]{cancel}
\usepackage[toc,page]{appendix}
\usepackage[colorlinks=true,citecolor=blue,linkcolor=magenta]{hyperref}

\usepackage{tikz}
\tikzset{every picture/.style=remember picture}

\renewcommand{\vec}[1]{\boldsymbol{#1}}  

\newcommand{\bs}{\textsf{BS}}

\def\be{\begin{equation}}
\def\ee{\end{equation}}
\def\bs{\begin{split}}
\def\e{\end{split}}
\def\ba{\begin{eqnarray}}
\def\bea{\begin{eqnarray}}

\def\tea{\end{eqnarray}}
\def\ea{\end{eqnarray}}
\def\eea{\end{eqnarray}}

\usepackage[utf8]{inputenc}

\usepackage{graphicx}
\usepackage{color}
\usepackage{dcolumn}
\usepackage{bm}
\usepackage{braket}
\usepackage{amsthm}
\usepackage{amssymb}
\usepackage{natbib}
\usepackage{siunitx}
\usepackage{booktabs}
\usepackage{mhchem}
\usepackage{threeparttable}
\usepackage{subfigure}
\usepackage{comment}
\usepackage[normalem]{ulem}
\usepackage{multirow}

\begin{document}

\preprint{APS/123-QED}


\title{Quantum Multiple Kernel Learning in Financial Classification Tasks}


\author{Shungo Miyabe}%
\email{Shungo.Miyabe@ibm.com}
\affiliation{%
 IBM Quantum, 19-21 Nihonbashi Hakozaki-cho, Chuo-ku, Tokyo, 103-8510, Japan
}%

\author{Brian Quanz}
\affiliation{IBM Quantum, Industry \& Technical Services, USA}

\author{Noriaki Shimada}%
\affiliation{%
 IBM Quantum, 19-21 Nihonbashi Hakozaki-cho, Chuo-ku, Tokyo, 103-8510, Japan
}%

\author{Abhijit Mitra}
\affiliation{IBM Quantum, Industry \& Technical Services, USA}

\author{Takahiro Yamamoto}%
\affiliation{%
 IBM Quantum, 19-21 Nihonbashi Hakozaki-cho, Chuo-ku, Tokyo, 103-8510, Japan
}
\author{Vladimir Rastunkov}
\affiliation{IBM Quantum, Industry \& Technical Services, USA}

\author{Dimitris Alevras}
\affiliation{IBM Quantum, Industry \& Technical Services, USA}

\author{Mekena Metcalf}
\email{mekena.metcalf@us.hsbc.com}
\affiliation{HSBC Holdings Plc., One Embarcadero Ctr. San Francisco, CA}

\author{Daniel J.M. King}
\affiliation{HSBC Holdings Plc., 8 Canada Square, London, UK}

\author{Mohammad Mamouei}
\affiliation{HSBC Holdings Plc., 8 Canada Square, London, UK}

\author{Matthew D. Jackson}
\affiliation{HSBC Holdings Plc., 8 Canada Square, London, UK}

\author{Martin Brown}
\affiliation{HSBC Holdings Plc., 8 Canada Square, London, UK}

\author{Philip Intallura}
\affiliation{HSBC Holdings Plc., 8 Canada Square, London, UK}

\author{Jae-Eun Park}
\affiliation{IBM Quantum, Industry \& Technical Services, USA}



\begin{abstract}Financial services is a prospect industry where unlocked near-term quantum utility could yield profitable potential, and, in particular, quantum machine learning algorithms could potentially benefit businesses by improving the quality of predictive models. Quantum kernel methods have demonstrated success in financial, binary classification tasks, like fraud detection, and avoid issues found in variational quantum machine learning approaches. However, choosing a suitable quantum kernel for a classical dataset remains a challenge. We propose a hybrid, quantum multiple kernel learning (QMKL) methodology that can improve classification quality over a single kernel approach. We test the robustness of QMKL on several financially relevant datasets using both fidelity and projected quantum kernel approaches.  We further demonstrate QMKL on quantum hardware using an error mitigation pipeline and show the benefits of QMKL in the large qubit regime.

\end{abstract}

\maketitle

\section{Introduction}

Quantum kernel-based methods are one of the major classes of approaches used for Quantum Machine Learning (QML) \cite{biamonte2017quantum}, and the quantum-enhanced Support Vector Machine (QSVM) \cite{havlivcek2019supervised} has become a "workhorse" in many QML applications \cite{Park2020practical,rebentrost2014quantum,peters2021machine,rastunkov2022boosting}.

One key application area for QML, and the focus of our experiments in this paper, is the financial services industry, with use cases encompassing fraud detection, default prediction, credit scoring, loan approval, directional forecasting of asset price movement, and buy/sell recommendations \cite{orus2019quantum,egger2020quantum,bouland2020prospects,herman2022survey}.
Quantum machine learning (QML), and specifically quantum kernel methods like QSVM, have demonstrated improvements when benchmarked against classical methods in fraud classification tasks \cite{grossi2022mixed, kyriienko2022unsupervised}. 
These studies serve as motivation to further explore finance applications and work toward improving practical performance when using quantum kernel methods in financial services.  
Discrimination quality, among other factors, significantly impacts both the business and customer by increasing true positives and reducing false positives, therefore, it can be beneficial to develop new methods to further improve classifiers.

However, achieving good results, in terms of accurate models, with kernel-based methods requires finding the right kernel for the given data \cite{scholkopf2002learning,kubler2021inductive,JMLR:v13:cortes12a}, and choosing a single, arbitrary kernel (as done in the prior work above) may not lead to the best fit for a given dataset.  Alternatively, quantum kernel alignment (QKA), employs a variational quantum circuit to learn a kernel which is optimized to maximize alignment (i.e., similarity) with a target kernel \cite{glick2021covariant}. This approach requires an expensive iterative procedure to optimize over the parameterized quantum circuits which also often suffers from barren plateaus \cite{thanasilp2022exponential,wang2021noise,Marrero2021,cerezo2021cost}, and learning an arbitrary kernel function like this can also lead to overfitting \cite{JMLR:v13:cortes12a}.

We propose an alternative approach for improved kernel-based QML that combines multiple quantum kernels to enhance model performance when the data is difficult to model using a single, arbitrary kernel, borrowing from a previous approach used for classical kernel-based machine learning referred to as multiple kernel learning (MKL) \cite{JMLR:v13:cortes12a}. Our quantum MKL approach uses a fixed set of quantum kernels that are linearly combined classically to create a new kernel that is better suited for a given dataset and task, and more robust for quantum-enhanced modeling using the resulting kernel. A classical solver determines the kernel weights, therefore, it enables learning a suitable quantum-enhanced kernel while avoiding the difficulties of optimizing a quantum circuit. We empirically study this approach and find it can also help overcome challenges of running quantum kernel methods on real quantum hardware (as has been previously observed \cite{thanasilp2022exponential}) by stabilizing classification performance when more features and corresponding encoding qubits are used.  Additionally, through numerical simulations, we show that this approach can provide benefit over classical methods and single kernel approaches for key financial datasets.  

We test our quantum multi-kernel learning method on multiple financially-related datasets including HSBC Digital Payment data. Both fidelity quantum kernel \cite{havlivcek2019supervised} and the more recent projected quantum kernel \cite{Huang2021} techniques were tested in simulation and demonstrated on quantum hardware.  Hardware implementation was enhanced using an error mitigation pipeline composed of randomized compiling to reduce coherent errors and pulse efficient transpilation to reduce the temporal overhead for cross-resonance gates for two-qubit unitary rotations. This pulse transpilation approach enabled us to scale our feature space up to 20 qubits on hardware, and, to our knowledge, is one of the larger quantum machine learning implementations demonstrated on real hardware.

The rest of the paper is organized as follows: the theoretical framework for quantum multiple-kernel learning is detailed in Section 1, the error mitigation pipeline is explained in Section 2, and in Section 3 detailed experiment results for the financial datasets are provided both for simulation and hardware execution.

\section{Theory}

\subsection{Quantum Kernels}


Following \cite{havlivcek2019supervised} we define a feature map on $n$-qubits as 
\begin{equation}
  \label{eq:feature_map}
  {\mathcal{U}}_{\Phi}(\vec{x})=U_{\Phi(\vec{x})}H^{{\otimes}n}
\end{equation}
where
\begin{equation}
  \label{eq:feature_map_diag_gate}
U_{\Phi(\vec{x})}=\exp\left(i\sum_{S\subseteq [n]}
\alpha_S \phi_S(\vec{x})\prod_{i\in S} P_i\right),
\end{equation}
which defines a data-point-dependent unitary transformation that is applied to an initial state $\rho_0$ (typically the $0$ state, which we use in our experiments) to get a transformed quantum state representation of a data point $\vec{x}$.   
Here $H$ is the Hadamard gate, $P_i \in \{ I, X, Y, Z \}$ are identity or Pauli matrices that correspond to different rotation types, and $\alpha_S$, typically restricted to a single shared value $\alpha_S = \alpha$, correspond to rotation scaling factors.  The subsets $S$ to use must also be specified and typically these include each single qubit along with an entangling pattern, e.g., to specify pairs of qubits to use, such as ``pairwise'' entanglement corresponding to odd and even pairs of qubits.  Finally $\phi_S(\vec{x})$ specifies how to use the feature values in each quantum operation; herein we follow a common approach in which a single feature value is assigned to each qubit with that feature value used for each single qubit operation for its corresponding qubit, and products of feature values for the corresponding qubits are used for each pairwise operation.  For example, a commonly used feature map for the above formulation can be specified with the sequence of Pauli strings ``Z-ZZ'' and a linear entanglement pattern, which results in the following definition:
\begin{equation}
  \label{eq:z_zz_fmap}
U_{\Phi(\vec{x})}=\exp\left(i \alpha (\sum_{i=1}^{n} x_i Z_i + \sum_{i=1}^{n-1} x_i x_{i+1} Z_i Z_{i+1}) \right),
\end{equation}
and we use this style of short-hand notation to describe feature maps of Eq. \ref{eq:feature_map_diag_gate} type going forward.

These feature maps correspond to inter-mixing entangling operations with different rotation operations where the angle of each rotation is given by one or more feature values, and is scaled by $\alpha$. Changing $\alpha$ affects how  similar in general resulting quantum states are for different data points, as smaller $\alpha$ leads to less change from the initial state for all data points, and thus higher similarity and less variance in similarity.  Therefore, since this can be viewed as controlling the ``width'' of a kernel function corresponding to the given feature map, i.e., a function measuring the similarity of two data points based on the feature map (defined below), $\alpha$ is also referred to as the kernel \emph{bandwidth} \cite{shaydulin2022importance}.  
 Changing $\alpha$ affects the complexity of a feature map as well as machine learning model over-fitting when using the feature map \cite{Park2020practical}, and also relates to trainability of models using a kernel based on the feature map \cite{shaydulin2022importance}. This point will be emphasized and discussed later in this section.
We note that in this scheme we use $n$ qubits to encode an $n$-dimensional data points. 

In quantum kernel methods, each input data point $\vec{x}_i$ is encoded into an $n$-qubit quantum state $\rho(\vec{x}_i)$ using a given feature map: 
\begin{equation}
  \label{eq:density_matrix}
  \rho(\vec{x}_i)=\mathcal{U}_{\Phi}(\vec{x}_i)\rho_0 \mathcal{U}_{\Phi}^\dag (\vec{x}_i),
\end{equation}
where again $\rho_0$ is some initial state.
For a given input data pair $\vec{x}$ and $\vec{x}'$ the fidelity kernel can be defined as
\begin{equation}
  \label{eq:fq-kernel}
  K^{\text{FQ}}(\vec{x},\vec{x}')=\text{Tr}\left[ \rho(\vec{x})\rho(\vec{x}') \right].
\end{equation}
One common way to compute this fidelity on quantum hardware is the compute-uncompute method \cite{havlivcek2019supervised}, which we use in our experiments as it requires no additional qubits beyond those required to compute a feature map.  

Finally, the projected quantum kernel \cite{Huang2021} is defined as
\begin{equation}
  \label{eq:pq-kernel}
  K^{\text{PQ}}(\vec{x},\vec{x}')=\exp
  \left( -\gamma\sum_{k=1}^n || \rho_k(\vec{x})-\rho_k(\vec{x}') ||_F^2 \right)
  ,
\end{equation}
where $||\cdot ||_F$ is the Frobenius norm and $\gamma$ is a positive hyperparameter, and $\rho_k(\vec{x})$ is the one-particle reduced density matrix (1-RDM) for qubit $k$ for the encoded state, i.e., $\rho_k(\vec{x}) = \text{Tr}_{j \neq k}\left[ \rho(\vec{x}) \right]$.  Projected quantum kernel values can be computed for a dataset more efficiently and with shallower circuits than fidelity quantum kernels since this approach amounts to computing a set of observables for each data point individually, with the intent to ``project'' the quantum state onto a reduced classical representation, and then subsequently computing the kernel value between each pair of data points via classical computation based on these classical representations. 

Note that a given kernel function $K(\vec{x},\vec{x}')$ results in a corresponding kernel matrix for a given pair of data samples, $S_1=\{ x_1,...,x_m \}$ and $S_2=\{ x_1',...,x_l' \}$, which corresponds to the kernel function evaluated between all pairs of data points in the two sets, and for simplicity we refer to such kernel matrices with notation $K$. Specifically, for the two sets of data points above, which could for example correspond to a test dataset and a train dataset, the $i^{th}$ row and $j^{th}$ column entry for kernel matrix $K$ is given by: $K_{ij} = K(\vec{x_i},\vec{x_j'})$, for $i=0, ..., m$, and $j=0, ..., l$.  We often refer to the kernel matrix for a single dataset or sample (such as for the set of training data) with the same notation as well, which is a symmetric matrix defined as above with $S_1 = S_2$ (the single data sample).  Furthermore, for simplicity may refer to both kernel functions and kernel matrices simply as kernels interchangeably  throughout the rest of the paper, where the meaning is clear given the context.  Note that valid kernel functions, as well as the corresponding kernel matrices for a given dataset, must be positive semi-definite \cite{scholkopf2002learning}.

\subsubsection{Exponential Concentration of Quantum Kernels}\label{exponential_concentration}

Before we introduce multiple kernel learning, we comment on the exponential concentration of kernel values that can occur with increasing number of qubits when computing quantum kernels on quantum hardware \cite{thanasilp2022exponential}.  As the number of features and thus qubits needed to compute a kernel increases, the difference between kernel values for different pairs of data points can become increasingly smaller thus requiring increasing shots to distinguish them.  This phenomenon can impede the training of any kernel-based methods and make it challenging to scale quantum kernel based methods to larger numbers of feature and qubits. A brief description here follows  \cite{thanasilp2022exponential}.
A quantity $X(\xi)$ that depends on variables $\xi$ is said to be probabilistically exponentially concentrated (in the number of qubits $n$) if 
\begin{equation}
  \label{eq:exp_conc}
  \text{Pr}_\xi\left[ |X(\xi)-\mu|\ge \delta \right]\le \frac{\beta^2}{\delta^2}, \beta\in O(1/b^n),
\end{equation}
for $b>1$.
Similarly, $X(\xi)$ is exponentially concentrated if
\begin{equation}
  \label{eq:exp_conc_var}
  \text{Var}_\xi\left[ X(\xi) \right] \in  O(1/b^n),
\end{equation}
for $b>1$ \cite{thanasilp2022exponential}.
Note that for quantum kernels, $\xi$ is a pair of input data, and thus, the probability in Eq.~\ref{eq:exp_conc} and the variance in Eq.~\ref{eq:exp_conc_var} is taken over all possible pairs of input data $\{\vec{x},\vec{x}'\}$.
Furthermore, since exponential concentration drives a kernel matrix $K$ towards a fixed kernel with diagonal elements $1$ and all off-diagonal elements $\mu$, we can simply plot the average of $|K^{\text{FQ}}(\vec{x},\vec{x}')-1/2^n|$ and $|K^{\text{PQ}}(\vec{x},\vec{x}')-1|$ for the two kernels that we consider in this report.
Both these averages and the variance will be computed and plotted in later sections and this point will be discussed further. We will show that exponential concentration can be avoided by tuning $\alpha$ in Eq.~\ref{eq:feature_map_diag_gate}, kernel bandwidth, or by using multiple kernel learning.

\subsection{Multiple kernel learning}\label{kernel_alignment}

In the multiple kernel learning (MKL) method investigated here we combine a set of kernels $K_i$ to construct a combined kernel $K$ that is optimized for a particular dataset and task in the following manner,
\begin{equation}
  \label{eq:mkl}
  {K = \sum_i^{N_K}  w_i K_i,}
\end{equation}
where $N_K$ is the number of kernels included, and $w_i \ge 0$ is the weight of a particular kernel $K_i$, which could be thought of as capturing the importance of that kernel in the combination.
$K_i$ is chosen from a set of predefined kernels $\mathcal{K}_S=\{ K_0, K_1, ... \}$.   Note that this form for the combined kernel guarantees that it is also a valid kernel (positive semi-definite), given each $K_i$ is a valid kernel.  
In this manner we use a fixed set of quantum kernels, and linearly combine them, obtaining optimal weights using a classical solver that maximizes kernel alignment with a target kernel for the task.

The kernel alignment score, a measure of similarity between two kernels given a data sample, is used here to determine the weights $w_i$ in Eq.~\ref{eq:mkl}.
This quantity for kernel matrices $K_1$ and $K_2$ (i.e., computed on the same data sample so having the same dimensions) is given by: 
\begin{equation}
  \label{eq:kernel alignment}
  {\hat{A}(K_1,K_2) = \frac{\left<K_1,K_2 \right>_F}{\sqrt{\left<K_1,K_1 \right>_F\left<K_2,K_2 \right>_F}}}
\end{equation}
where $\left<K_1,K_2 \right>_F=\sum_{i,j=1}^mK_1(x_i,x_j)K_2(x_i,x_j)$ is an inner product between kernel matrices given a sample $S=\{ x_1,...,x_m \}$.
More concretely, given $K$ from (\ref{eq:mkl}) and a target kernel matrix $K_y$, we maximize $\hat{A}(K,K_y)$ with respect to $w_i$ to achieve optimal alignment between $K$ and $K_y$,
\begin{equation}
  \label{eq:ka-optimization}
  \max_{w_i} A\left( K,K_y \right) \quad \textrm{s.t.} \quad Tr(K)=1, \quad w_i \ge 0,
\end{equation}
where $i,j^\text{th}$ element of $K_y$ for a classification task with corresponding labels $y_i$ for $i = 1, ..., m,$ is defined as
\begin{equation}
  \label{eq:target-kernel}
  (K_y)_{ij} = 
  \begin{cases}
  1, &\text{if } y_i = y_j \\
  0, &\text{otherwise}.
  \end{cases}
\end{equation}

We examine three strategies to optimize $w_i$: (1) kernel-target alignment with semidefinite programming (SDP) \cite{JMLR:v5:Lanckriet}, (2) centered alignment \cite{JMLR:v13:cortes12a} and (3) iterative projection-based alignment. 

\subsubsection{Kernel-target Alignment with SDP}
A maximally aligned kernel matrix $K$ can be determined by solving the following SDP problem:
\begin{align}
    \max_{K} \quad &\null A(K, K_y) \label{eq:alignment_SDP_base} \\
    \text{subject to} \quad &\null K \in \mathcal{K}, \nonumber \\
    &\null \text{Tr}(K) \le 1 \nonumber
\end{align}
where $\mathcal{K}$ denotes some class of positive semidefinite kernel matrices. If $K$ is a linear combination of fixed kernel matrices as Eq.~\ref{eq:mkl}, Eq.~\ref{eq:alignment_SDP_base} can be written in the standard form of SDP:
\begin{align}
    \max_{A, w_i} \quad &\null \big\langle \sum^{N_K}_{i=1}w_iK_i, K_y \big\rangle_F \label{eq:alignment_SDP}\\
    \text{subject to} \quad &\null \text{Tr}(A) \le 1, \nonumber \\
    &\null \begin{pmatrix}
            A & \sum^{N_K}_{i=1}w_iK^T_i \nonumber \\
            \sum^{N_K}_{i=1}w_iK_i & I_m \nonumber \\
        \end{pmatrix} \succeq 0, \nonumber \\
    &\null \sum^{N_K}_{i=1}w_iK_i \succeq 0 \nonumber
\end{align}
where $I_m$ is the identity matrix of dimension $m$, the number of data points. If $\bm{w} \ge 0$ and $K_i \succeq 0$, Eq.~\ref{eq:alignment_SDP} can be reduced to the following quadratically constrained quadratic program (QCQP):
\begin{align}
    \max_{\bm{w}} \quad &\null \bm{w}^T \bm{q} \label{eq:alignment_QCQP} \\
    \text{subject to} \quad &\null \bm{w}^T \bm{S} \bm{w} \le 1, \nonumber \\
    \quad &\null \bm{w} \ge \bm{0} \nonumber
\end{align}
where $q_i = \langle K_i, K_y \rangle_F$ and $S_{i, j} = \langle K_i, K_j \rangle_F$. Kernel-target alignment with SDP actually solves Eq.~\ref{eq:alignment_QCQP}.

\subsubsection{Centered Alignment}
The centered kernel matrix $K_c$ is defined as
\begin{align}
    K^c_i = \biggl[ \bm{I}_m - \frac{\bm{1}\bm{1}^T}{m} \biggr] K_i \biggl[ \bm{I}_m - \frac{\bm{1}\bm{1}^T}{m} \biggr]
\end{align}
where $\bm{1} \in \mathbb{R}^{m \times 1}$ denotes the vector with all elements equal to one. This corresponds to the kernel computed after centering each data point in the feature space - that is, subtracting the mean of the data points in the feature space from each data point.  After centering, it was previously shown that the alignment score often better correlates with kernel method  generalization performance \cite{JMLR:v13:cortes12a}. Centered alignment optimizes $w_i$ by solving the following optimization problem:
\begin{align}
    \max_{\bm{w}} \quad &\null A(K_c, K_y^c) \label{eq:alignment_centered_base} \\
    \text{subject to} \quad &\null \| \bm{w} \|^2 = 1, \nonumber \\
    &\null \bm{w} \ge \bm{0} \nonumber
\end{align}
where $K_c = \sum^{N_K}_{i=1} w_i K^c_i$. Optimal weights $\bm{w}^*$ of Eq.~\ref{eq:alignment_centered_base} can be written as
\begin{align}
    \bm{w}^* = \underset{\bm{w} \in \mathcal{M}} {\operatorname{argmax}} \frac{\bm{w}^T \bm{a} \bm{a}^T\bm{w}}{\bm{w}^T \bm{M} \bm{w}}
\end{align}
where $\mathcal{M} = \{ \| \bm{w} \|^2 = 1 \cup \bm{w} \ge \bm{0}\}$, $a_i = \langle K^c_i, K_y^c \rangle_F$ and $M_{i, j} = \langle K^c_i, K^c_j \rangle$. Let $\bm{v}^*$ the solution of the following optimization problem:
\begin{align}
    \min_{\bm{v} \ge \bm{0}} \quad &\null \bm{v}^T \bm{M} \bm{v} - 2 \bm{v}^T \bm{a} \label{eq:alignment_centered}
\end{align}
Then, $\bm{w}^*$ can be obtained as $\bm{w}^* = \bm{v}^*/ \| \bm{v}^* \|$. Centered alignment actually solves Eq.~\ref{eq:alignment_centered}.

\subsubsection{Alignment through projection}

An alternative approach we propose to carry out target-kernel-alignment is through a residual after matrix projection
\begin{equation}
    K_y' =  \frac{1}{2} \left[ K_y - \hat{K_y} (K^T \hat{K_y}) \right].
    \label{eq:matrix_proj}
\end{equation}
where $K\in \mathcal{K}_S$ and $\hat{K_y}$ is the normalized $K_y$. 
In the above equation, we project $K$ onto $K_y$ and subtract that component from $K_y$ to obtain the residual.
This expression serves two purposes. First, the norm of $K_y'$ ($|K_y'|$) will be used as a criteria to truncate the summation in Eq.~\ref{eq:mkl}. As $K_i$ are added to the expansion in an iterative fashion, if the computed norm increases by addition or goes below a chosen threshold, the expansion will be truncated. In addition, the norm will also be used to determine $w_i$. This corresponds to giving more importance to the kernels that contribute more to a better alignment with the data (target kernel).

Now we describe the steps taken to choose the kernels and their weights.
\begin{enumerate}
    \item Starting with $K_y$ choose the kernel $K$ from $\mathcal{K}_S$ that has the shortest distance to $K_y$, $\left| K - K_y \right|$.
    \item Subtract the components of $K$ from $K_y$ using Eq.~\ref{eq:matrix_proj} and obtain $K_y'$.
    \item Compute $\left|K_y'\right|$, and compare it to the norm before the subtraction.
    \item If the current norm is less than the norm of previous iteration, add another kernel by iterating steps 2-4 using another $K$. Note that in the next iteration $K_y'$ is used in place of $K_y$ in step 3 to ensure that multiple kernel contribution is evaluated.
    \item Terminate iteration if the current norm is larger than the previous norm or if it is below a threshold, and we normalize the weights at the end of iterations.
\end{enumerate}

With this approach, a suitable kernel is iteratively constructed that is well-aligned with the target kernel, which includes those kernels that provide the biggest individual improvement in the alignment at each step, while generally avoiding overly redundant or overlapping kernels.

\section{Error Mitigation}

Randomized compiling and pulse efficient transpilation were employed to reduce the overhead of stochastic and coherent gate errors. Measurement error mitigation is conducted by computing the calibration matrix on the $2^N$ basis states and fitting subsequent experimental measurements with this matrix.

Coherent errors can arise from cross-talk, unwanted qubit correlations, or imperfect control of unitary gate implementations like the arbitrary $SU(2)$ rotations required for many near term algorithms. Error mitigation and  error correction methods are designed to resolve stochastic, incoherent errors, therefore, it is desirable to have incoherent errors rather than coherent errors on quantum computers. Randomized compiling transforms coherent errors to incoherent errors through the introduction of twirling operators. These twirling operators consist of 'easy' gates, (e.g. Pauli operators) to implement on hardware that sandwich hard gates (e.g. arbitrary rotations), and the noise is tailored by averaging over independent random sequences~\cite{wallman2016twirl}. We used a total of 16 independent random Pauli twirling sequences for the basis gates $U_Z(\theta)$, $\sqrt{X}$, $U_{ZZ}(\theta)$ to reduce coherent error in the quantum machine learning experiments. 

\begin{figure}
\centering
    \includegraphics[width=3.25in]{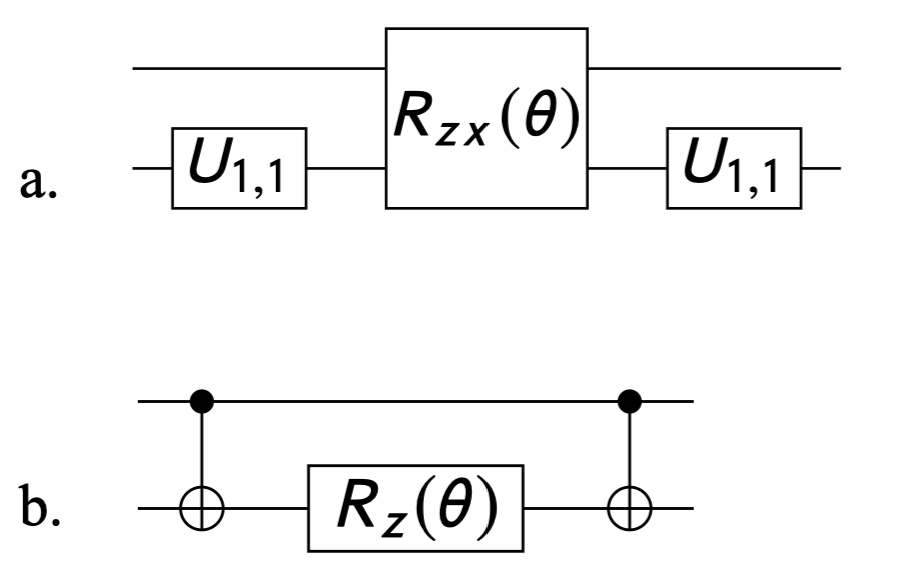}
    \caption{\label{Fig:zx_gate} (a) Circuit diagram of $U_{ZZ}(\theta)$ operator decomposed into single qubit gates and $ZX(\theta)$ gate. (b) Standard $U_{ZZ}(\theta)$ into CNOTs and $R_z(\theta)$ rotations.
    }
\end{figure}

Two qubit interaction sequences contribute to much of the error in quantum circuit execution as the pulse time to implement these unitaries is considerably longer than single qubit gate times. Circuit transpilation using the native gate set in the specified quantum processor can reduce the time it takes to execute SU(4) unitaries rather than using the standard circuit transpilation routines. The cross resonance gate is described by a unitary rotation in the ZX basis,
\begin{equation}
U_{ZX}(\theta) = e^{-i\theta ZX},
\end{equation}
with additional tones to suppress the undesired $I\otimes Y$ interaction. The universal CNOT gate is implemented by choosing the unitary $U_{ZX}(\pi/2)$, yet  more efficient transpilation is achievable using arbitrary rotations $\theta$ that permit scaling of the pulse area. Ref~\cite{Earnest2021pulse} demonstrated the pulse duration of the two-qubit $U_{ZZ}(\theta)$ unitary using the gate sequence in Fig~\ref{Fig:zx_gate}(a) is reduced to near a third of the cycle time compared to the standard double-CNOT implementation Fig~\ref{Fig:zx_gate}(b), so we compile using the gate sequence in Fig~\ref{Fig:zx_gate}(a).  Pulse efficient transpilation is extended to general SU(4) unitaries using the Cartan decomposition, and other canonical two-qubits gates are implemented by a basis change preceding the $U_{ZX}(\theta)$ rotation. 

\section{Numerical Results}

This section is divided into two parts. First, we will present our results on a simulator to describe the behavior of quantum MKL models under ideal conditions. We will compare performance of MKL models built using different types of quantum kernels, namely fidelity, projected and hybrid quantum and classical kernels. We will also measure the benefit of quantum MKL models against classical MKL and single kernel approaches. We will show that for the HSBC digital payment fraud and Bank Marketing datasets, quantum MKL offers the best performance. For the German Credit dataset, the performance of quantum MKL is equivalent to other approaches considered. We will discuss exponential concentration of kernel values with increasing number of qubits that can inhibit the trainability of models, and we will show that MKL can be used to mitigate such problems.

In the second part of this section we will discuss our results on actual quantum hardware using an IBM quantum computer. Here we will show that the error mitigation and suppression pipeline implemented here is effective in building quantum kernels used for classification tasks. We have evaluated the classification performance of quantum MKL models computed on the hardware against single kernel approaches. We will show that compared to single kernel models, quantum MKL offers better consistency in its performance on hardware.

\subsection{Results: Simulator}

In this work we consider several datasets to ensure consistency of the model. HSBC digital payment fraud dataset, German Credit data \cite{misc_statlog_(german_credit_data)_144} and Bank Marketing data \cite{MORO201422} are considered. 
The German Credit dataset classifies people characterized by a set of features as good or bad credit risks. It has 20 features and 1,000 instances.
The Bank Marketing dataset is another classification data that was collected during direct marketing campaigns of a Portuguese banking institution. The goal is to predict if the client will subscribe a term deposit. There are 16 features and over 45,000 instances.
For all datasets, we use 400 data points for evaluation. Note that the total number of data points for all datasets are much more than 400.
Thus, to ensure robustness of our analysis, we evaluated our models on 20 randomly drawn samples. We will refer to these 20 samples throughout the manuscript.

Following the best practice, the data is split into training, validation and testing datasets. For each of the 400 data points, $33\%$ was used as a test data. For the remaining $67\%$, the 4-fold cross validation was carried out for hyperparameter optimization. 

The features were standardized by subtracting the mean and scaling to unit variance. Then, the feature dimension was reduced using principal component analysis. Feature dimensions between 4 and 20 were used in this study. Finally, each feature was scaled to 0-2 range in order to restrict rotation angles used in the quantum feature maps to a reasonable range as well for default scaling factors of $1$. 

As mentioned, we employed two types of quantum kernels with our quantum MKL approaches, fidelity (FQ-MKL) and projected (PQ-MKL).
The classical kernels (C-MKL) used to evaluate the quantum models were built using radial basis function (RBF) kernels with varying kernel bandwidths ($\gamma$ hyper parameter values). The hybrid models used both quantum and classical kernels (CQ-MKL), where all fidelity, projected and RBF kernels were included. 

As mentioned in Sec.~\ref{kernel_alignment}, several strategies were taken to optimize the kernel weights ($w_i$ in Eq.~\ref{eq:mkl}). In addition to using the averaged weights (AVE), we use the weights optimized using kernel-target alignment with SDP (SDP) \cite{JMLR:v5:Lanckriet}, centered alignment (CENT) \cite{JMLR:v13:cortes12a}, and alignment through projection (PROJ). 

Finally, we note the use of kernels with relatively high values of $\alpha$ ($\sim 20$) in our quantum MKL models (see Eq.~\ref{eq:feature_map_diag_gate}). Their presence in MKL framework seems to help improve model performance. We tabulate the parameters of quantum kernels used in this work in table.~\ref{Table:kernel_params}. Note that the \textquote{linear} entanglement scheme is used throughout.
\begin{table}[!ht]
    \centering
    \begin{tabular}{|l|l|l|}
    \hline
        $P_i$ & $\alpha$ & reps \\ \hline
        Z & 1.4, 2, 14, 20 & 1 \\ \hline
        XZ & 0.4, 4.0  & 2 \\ \hline
        X-ZY & 0.6, 6.0 & 2 \\ \hline
        Y-XX & 0.6, 6.0 & 2 \\ \hline
        Y-XY & 1.4, 10 & 1 \\ \hline
        Y-XZ & 0.8, 8.0 & 2 \\ \hline
        Y-YX & 0.2, 2.0, 1.6, 1.6 & 1 \\ \hline
        Y-YZ & 1.2, 12 & 1 \\ \hline
        Y-ZX & 2.0, 20 & 1 \\ \hline
        Z-XX & 1.0, 10 & 1 \\ \hline
        Z-ZZ & 2.0, 20 & 1 \\ \hline
    \end{tabular}
    \caption{\label{Table:kernel_params}
    Parameters of quantum kernels used in this work are tabulated. See Eq.~\ref{eq:feature_map_diag_gate} for the definition of $P_i$ and $\alpha$.}
\end{table}

\begin{figure*}
\centering
    \includegraphics[width=\textwidth]{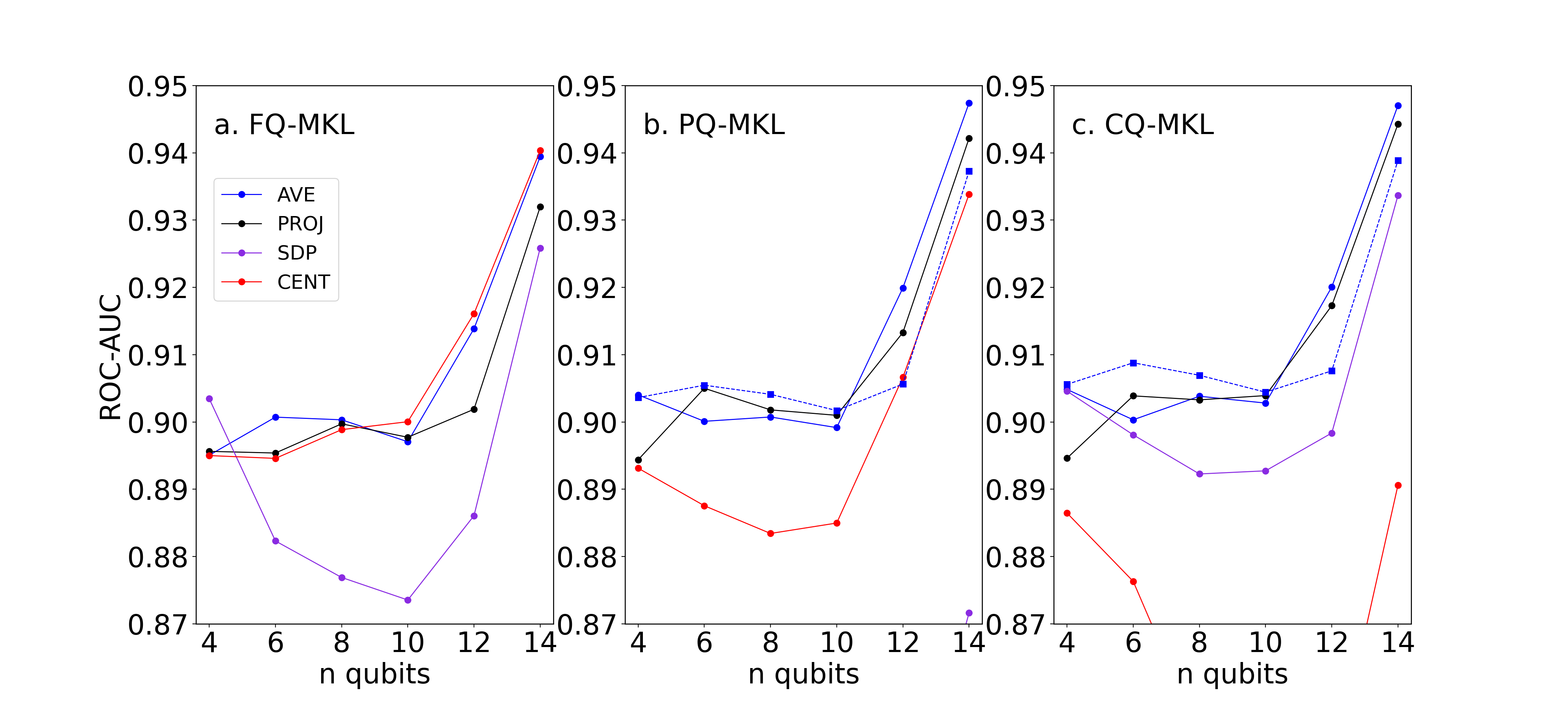}
    \caption{\label{Fig:qmkl_real_data} 
    The average test ROC-AUC are plotted for MKL models built using the kernel-target alignment schemes discussed in Sec.~\ref{kernel_alignment}, AVE, SDP, CENT and PROJ. The result is averaged over 20 samples. (a) Fidelity (FQ-MKL), (b) projected (PQ-MKL) and (c) hybrid (CQ-MKL) kernels were employed in our analysis. ROC-AUC is plotted for different number of qubits, n qubits, or feature dimensions. HSBC digital payment fraud dataset is used.
    }
\end{figure*}
\begin{figure*}
\centering
    \includegraphics[width=\textwidth]{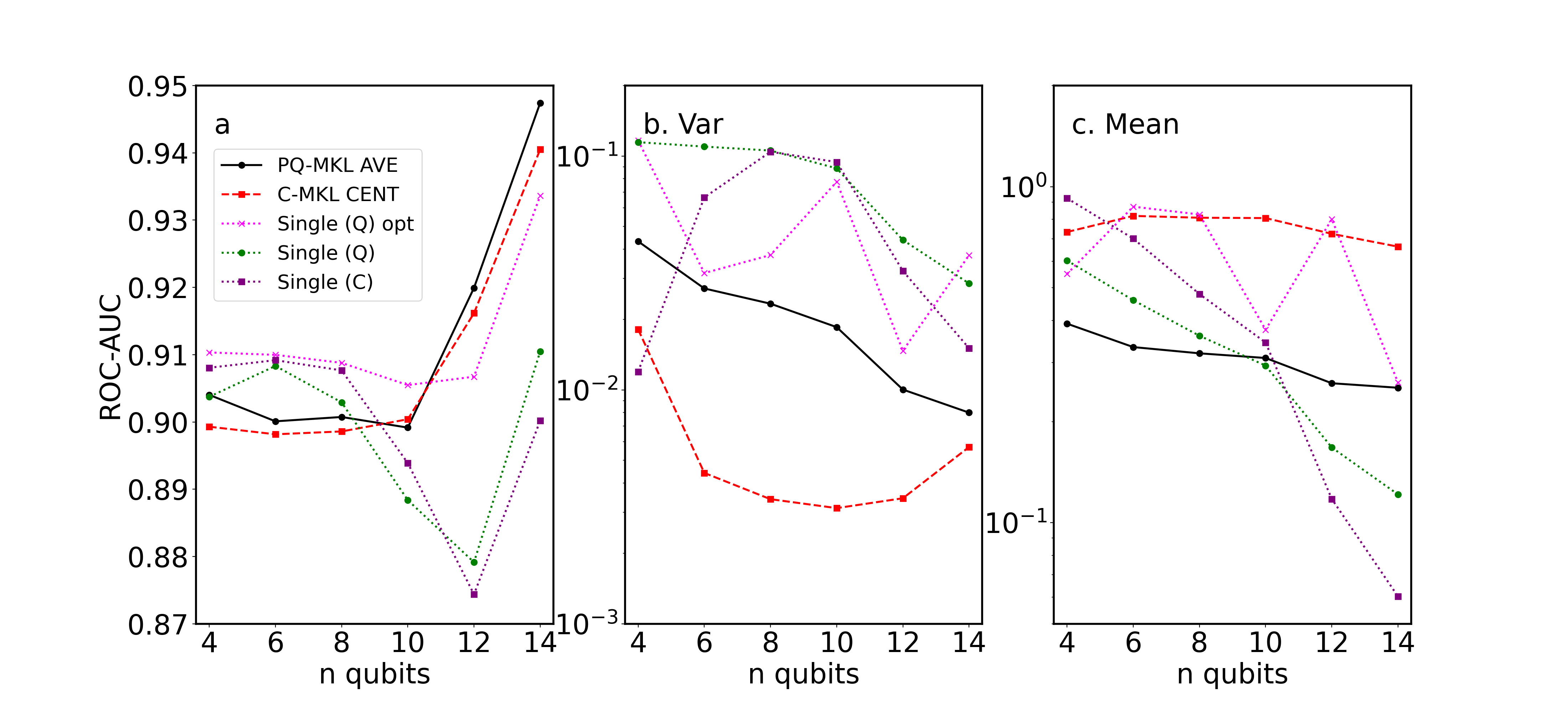}
    \caption{\label{Fig:mkl_real_data} 
    (a) ROC-AUC of a quantum model, PQ-MKL AVE, is compared with a classical model (C-MKL CENT) and representative single learner models for the HSBC digital payment fraud dataset. For the single learners, Single (Q) employed a ZZ-feature map with $\alpha=0.4$, repetition of $1$ and the linear entanglement scheme, and a radial basis function (RBF) with default length is used in Single (C). Single (Q) Opt is obtained by tuning the hyper-parameter with the parameters of quantum MKL. (b) Variance and (c) mean of kernel matrix used to build the SVM models in (a) are plotted. 
    }
\end{figure*}

\subsubsection{HSBC digital payment fraud dataset}

We first turn to our results on the HSBC digital payment fraud dataset. 
In Fig.~\ref{Fig:qmkl_real_data} we compare the performance, measured in ROC-AUC, of various target-kernel alignment approaches discussed in Sec.~\ref{kernel_alignment}. Note that the results are averaged over the 20 samples.
The advantage of the alignment, more specifically centered alignment (CENT), is observed only in the fidelity quantum kernel case (Fig.~\ref{Fig:qmkl_real_data}a). For the projected and hybrid models the performance is best when average weights (AVE) are used (Fig.~\ref{Fig:qmkl_real_data}b and Fig.~\ref{Fig:qmkl_real_data}c). We see that the best performing model across all parameters is PQ-MKL AVE.

In table.~\ref{Table:real_data_counts} we compare the performance of MKL models across the 20 samples by tabulating the number of times each MKL model gives the best ROC-AUC performance. We limited our comparison to the best performing models in each kernel type, FQ-MKL, PQ-MKL and CQ-MKL, and the best classical model. In this view, we can clearly see the advantage of PQ-MKL AVE, giving the best performance for all dimensions. FQ-MKL CENT and CQ-MKL AVE were competitive at dimensions 6 and 14, respectively. We note that all quantum models outperformed the best classical model in this view.

\begin{table}[!htbp]
\centering
\begin{tabular}{*5c}
\toprule
Kernel &  \multicolumn{3}{c}{Dimension} \\
\midrule
{} & 6 & 10 & 14  \\
FQ-MKL CENT & 6 & 4 & 1  \\
PQ-MKL AVE & 6 & 6 & 8  \\
CQ-MKL AVE & 5 & 6 & 8  \\
C-MKL CENT & 3 & 4 & 3  \\
\bottomrule
\end{tabular}
\caption{\label{Table:real_data_counts}
The performance of MKL models across the 20 samples is tabulated for the HSBC digital payment fraud dataset. Here the number of times each MKL model gives the best ROC-AUC over the samples is shown.}
\end{table}

In Fig.~\ref{Fig:mkl_real_data} we demonstrate the trainability of quantum MKL models with varying feature dimensions. 
We first note that the performance of SVM models are directly linked to the variance and the mean of kernel matrix elements plotted in Fig.~\ref{Fig:mkl_real_data}b and c, respectively. Note that they all relate back to the exponential concentration discussed in Sec.~\ref{exponential_concentration}.
A point to mention is that we previously, in Sec.~\ref{exponential_concentration}, suggested to plot the average of $|K^{\text{FQ}}(\vec{x},\vec{x}')-1/2^n|$ and $|K^{\text{PQ}}(\vec{x},\vec{x}')-1|$. However, since we prefer to plot various results in a single plot, we simply plot the mean value of kernel matrix elements for simplicity.

To proceed with the discussion, when there is concentration in the mean value with increasing feature dimension, the model trainability is inhibited.
This is demonstrated by the drop in ROC-AUC (Fig.~\ref{Fig:mkl_real_data}a) of the single learner methods, Single (Q) and Single (C). Here the hyperparameter of these kernels was not optimized.
To remedy this Shaydulin and Wild \cite{PhysRevA.106.042407} suggested to tune $\alpha$.
In our case, this is shown by Single (Q) Opt giving the best performance at dimensions $4\sim10$. Note that the mean kernel values remain consistent with increasing dimensions for this method. The fidelity kernel is used in Single (Q) Opt.
Another point the we emphasize is that MKL can also be used to overcome the exponential concentration. This is demonstrated by the superior performance of PQ-MKL AVE at higher dimensions ($>10$) and its mean kernel values, which remain consistent.

\subsubsection{Public Datasets}

\begin{figure*}
\centering
    \includegraphics[width=\textwidth]{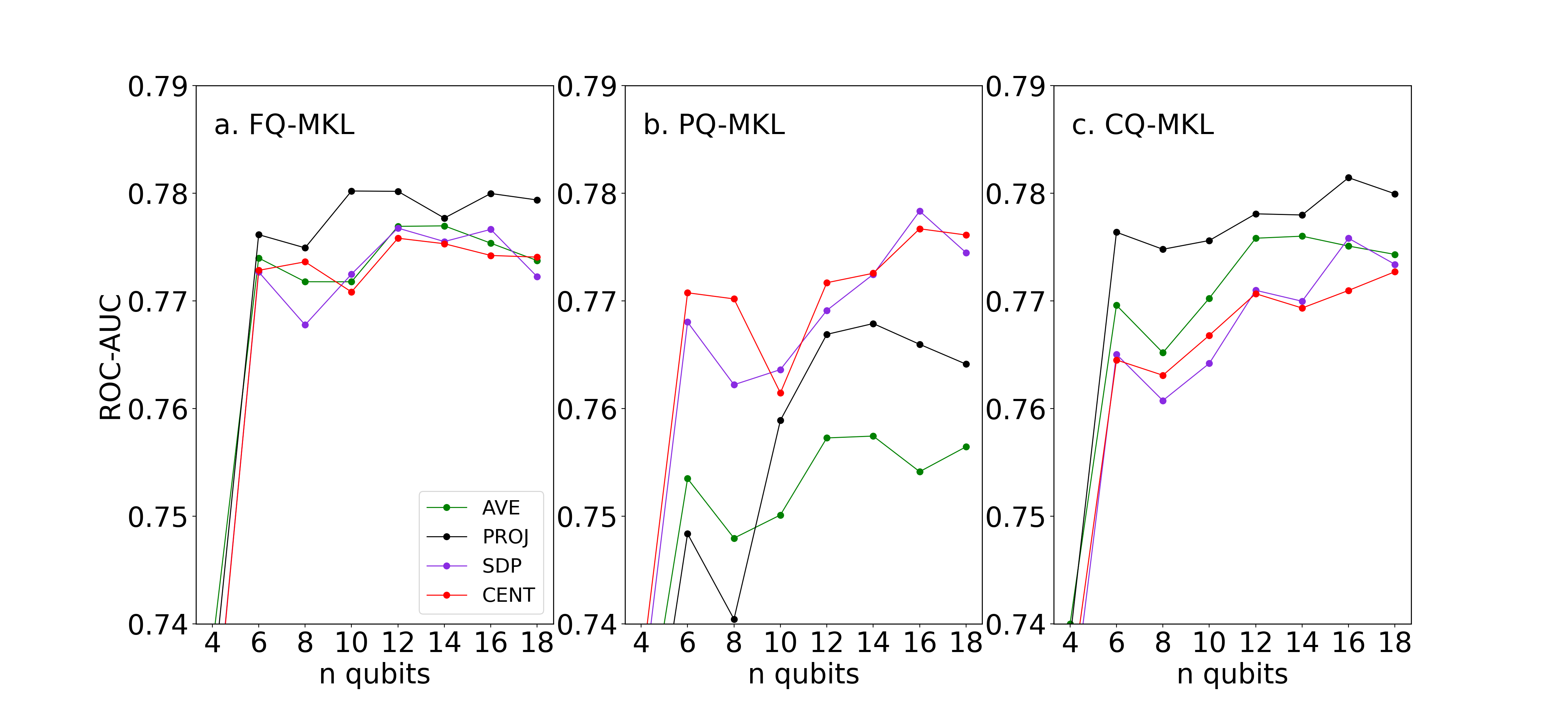}
    \caption{\label{Fig:qmkl_german_numeric} 
     We plot the average test ROC-AUC for the German numeric dataset. The results are plotted for different number of qubits, n qubits, or feature dimensions. Kernel weights $w_i$ are optimized in four different ways (AVE, PROJ, SDP and CENT). All result are averaged over the 20 samples. The results of (a) fidelity (FQ-MKL), (b) projected (PQ-MKL) and (c) hybrid (CQ-MKL) kernels are plotted.
    }
\end{figure*}

\begin{figure*}
\centering
    \includegraphics[width=\textwidth]{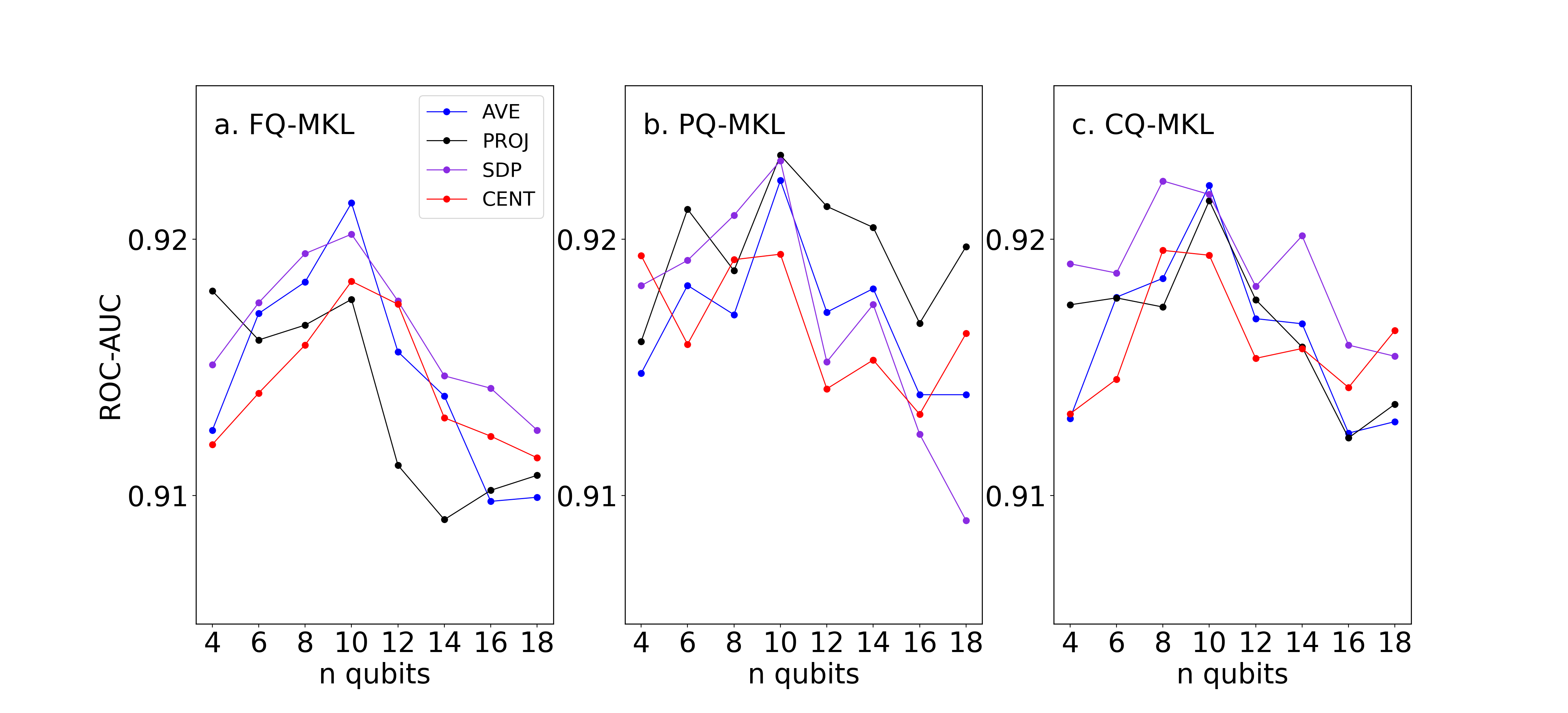}
    \caption{\label{Fig:qmkl_bank_marketing} 
     We plot the average test ROC-AUC for the Bank Marketing dataset. The results are plotted for different number of qubits, n qubits, or feature dimensions. Kernel weights $w_i$ are optimized in four different ways (AVE, PROJ, SDP and CENT). All result are averaged over the 20 samples. The results of (a) fidelity (FQ-MKL), (b) projected (PQ-MKL) and (c) hybrid (CQ-MKL) kernels are plotted.
    }
\end{figure*}

Now we turn to our results on the German Credit and Bank Marketing datasets.
In Fig.~\ref{Fig:qmkl_german_numeric} and \ref{Fig:qmkl_bank_marketing} we compare the performance of the target-kernel alignment approaches discussed in Sec.~\ref{kernel_alignment}.
This optimization effort seems to be highly effective for these datasets. 
For the German Credit dataset, alignment through projection (PROJ) is found to be the most effective approach for the fidelity (Fig.~\ref{Fig:qmkl_german_numeric}a) and the hybrid (Fig.~\ref{Fig:qmkl_german_numeric}c) cases. For the projected kernels (Fig.~\ref{Fig:qmkl_german_numeric}b) centered alignment (CENT) gives the best performance.
Similarly for the Bank Marketing dataset, PROJ is found to be most effective for FQ-MKL and CQ-MKL (Fig.~\ref{Fig:qmkl_bank_marketing}a and c), while CENT gives the best performance for PQ-MKL (Fig.~\ref{Fig:qmkl_bank_marketing}b).
For both datasets, PROJ optimization scheme provides the best performance in terms of ROC-AUC. Note that FQ-MKL gives the best performance on the German Credit dataset, while PQ-MKL is the best model for the Bank Marketing dataset.

In table.~\ref{Table:german_numeric_counts} we compare the performance of MKL models across the 20 samples for the German Credit data. We limited our comparison to the best performing quantum models in each category, and the best performing classical model, C-MKL AVE.
We can see that quantum models are comparable to the classical model in this view, giving equal or better performance in all feature dimensions. 
Similarly, in Table.~\ref{Table:bank_marketing_counts} we measure the same performance for the Bank Marketing dataset. Here we can clear see the advantage of PQ-MKL PROJ, giving the best performance across all dimensions.

\begin{table}[!htbp]
\centering
\begin{tabular}{*5c}
\toprule
Kernel &  \multicolumn{4}{c}{Dimension} \\
\midrule
{}  & 6 & 10 & 14 & 18  \\
FQ-MKL PROJ & 7 & 4 & 3 & 7 \\
PQ-MKL PROJ & 5 & 4 & 7 & 6 \\
CQ-MKL PROJ & 5 & 7 & 3 & 2 \\
C-MKL AVE   & 3 & 5 & 7 & 5 \\
\bottomrule
\end{tabular}
\caption{\label{Table:german_numeric_counts}
Same values shown in Table.~\ref{Table:real_data_counts} are tabulated for the Bank Marketing dataset.}
\end{table}

\begin{table}[!htbp]
\centering
\begin{tabular}{*5c}
\toprule
Kernel &  \multicolumn{4}{c}{Dimension} \\
\midrule
{} & 6 & 10 & 14 & 18  \\
FQ-MKL SDP & 5 & 4 & 3 & 3 \\
PQ-MKL PROJ & 11 & 8 & 6 & 10 \\
CQ-MKL SDP & 3 & 5 & 5 & 3 \\
C-MKL SDP  & 1 & 3 & 6 & 4 \\
\bottomrule
\end{tabular}
\caption{\label{Table:bank_marketing_counts}
Same values shown in Table.~\ref{Table:real_data_counts} are tabulated for the Bank Marketing dataset.
}
\end{table}

\begin{figure*}
\centering
    \includegraphics[width=\textwidth]{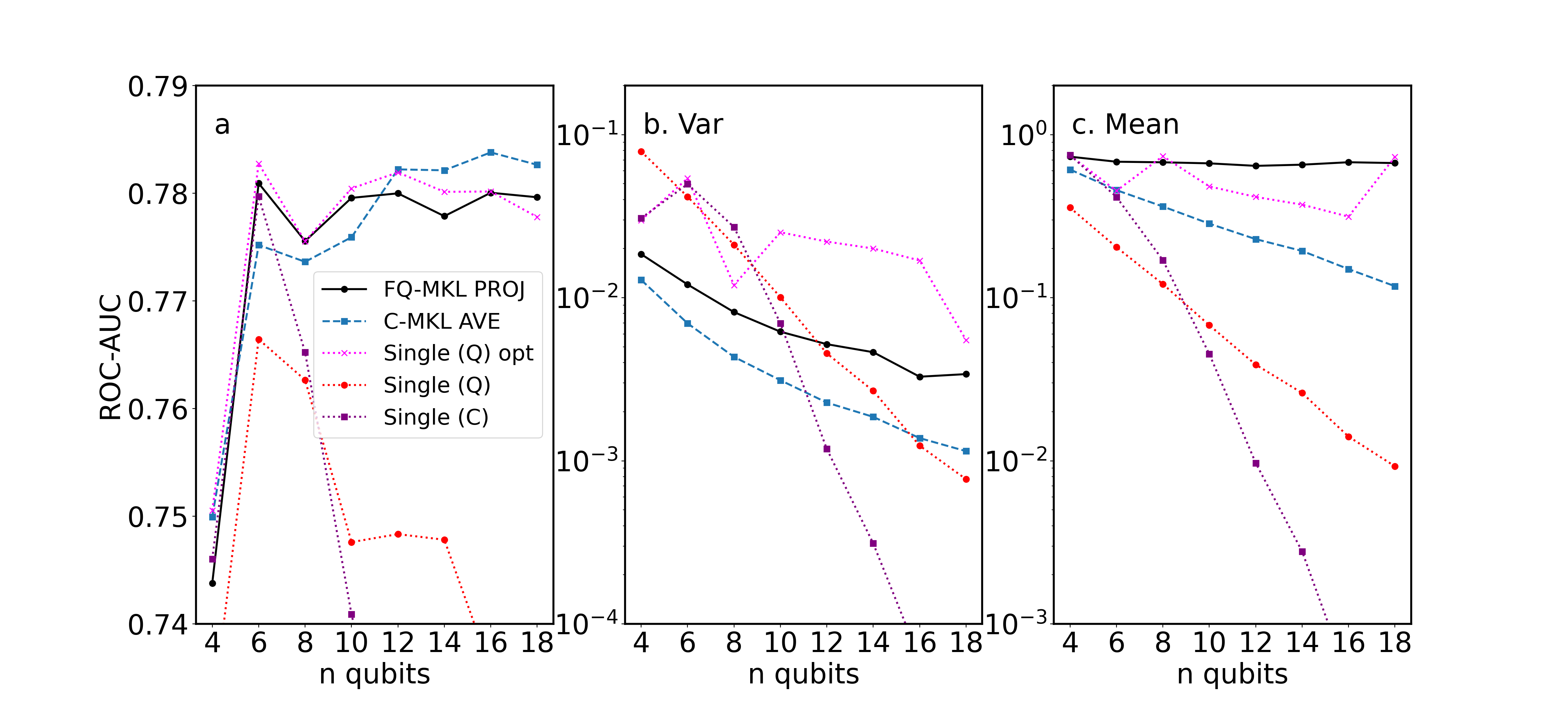}
    \caption{\label{Fig:mkl_german_numeric} 
    (a) ROC-AUC of a quantum model, FQ-MKL PROJ, is compared with a classical model (C-MKL AVE) and representative single learner models for the German Credit data. The description of single learners are equivalent to Fig.~\ref{Fig:mkl_real_data}. (b) Variance and (c) mean of kernel matrix used to build the SVM models in (a) are plotted.
    }
\end{figure*}

\begin{figure*}
\centering
    \includegraphics[width=\textwidth]{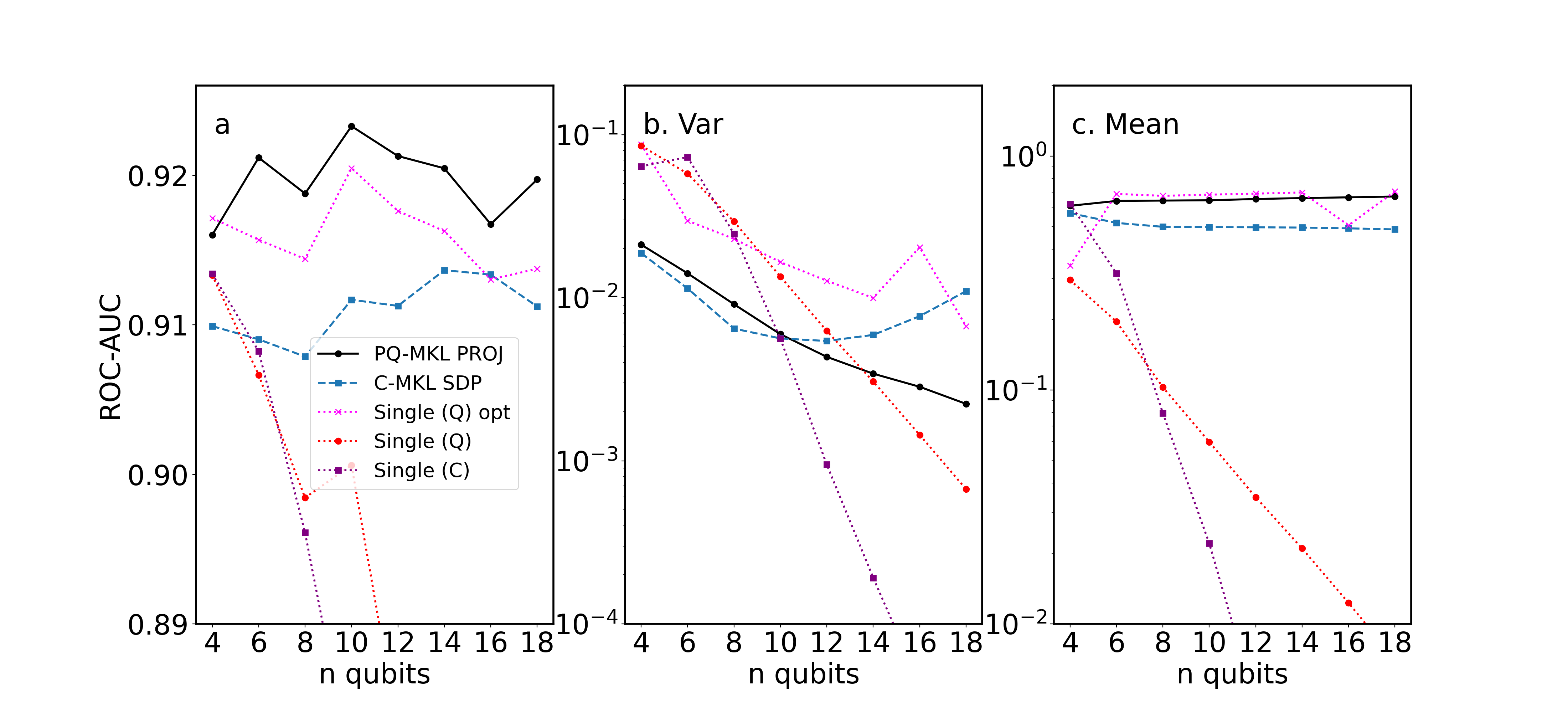}
    \caption{\label{Fig:mkl_bank_marketing} 
    (a) ROC-AUC of a quantum model, PQ-MKL PROJ, is compared with a classical model (C-MKL SDP) and representative single learner models for the Bank Marketing data. The description of single learners are equivalent to Fig.~\ref{Fig:mkl_real_data}. (b) Variance and (c) mean of kernel matrix used to build the SVM models in (a) are plotted.
    }
\end{figure*}

Relating back, once again, to Sec.~\ref{exponential_concentration}, the trainability and exponential concentration of quantum MKL models are investigated for the two datasets.
In both cases quantum MKL models are demonstrated to avoid the concentration that inhibit kernel trainability and give better or competitive performance compared to classical and single learner approaches considered here.
For the German Credit data, FQ-MKL PROJ gives good performance, measured in ROC-AUC, but we see that Single (Q) Opt and C-MKL AVE are better (Fig.~\ref{Fig:mkl_german_numeric}a). Although the performance of FQ-MKL is inferior to these approaches, we see that mean and variance of its kernel elements are consistent over varying dimensions, demonstrating its robustness (Fig.~\ref{Fig:mkl_german_numeric}b and c).
For the Bank Marketing data, PQ-MKL PROJ gives the best performance compared to all the other approaches (Fig.~\ref{Fig:mkl_bank_marketing}a).  
This approach is also demonstrated to be robust against kernel concentration (see Fig.~\ref{Fig:mkl_bank_marketing}b and c)
The untrainability of kernels without hyperparameter optimization is demonstrated by the results of Single (Q) and Single (C) for both datasets. The SVM models built using these kernels have poor ROC-AUC that drops rapidly with increasing dimensions. Their mean kernel values and variance also drop.
Hyperparameter optimization is shown to mitigate this concentration problem. In fact, for the German Credit dataset Single (Q) opt gives the best performance for dimensions less than $12$.

\subsection{Results: Hardware}

\begin{figure*}
\centering
    \includegraphics[width=\textwidth]{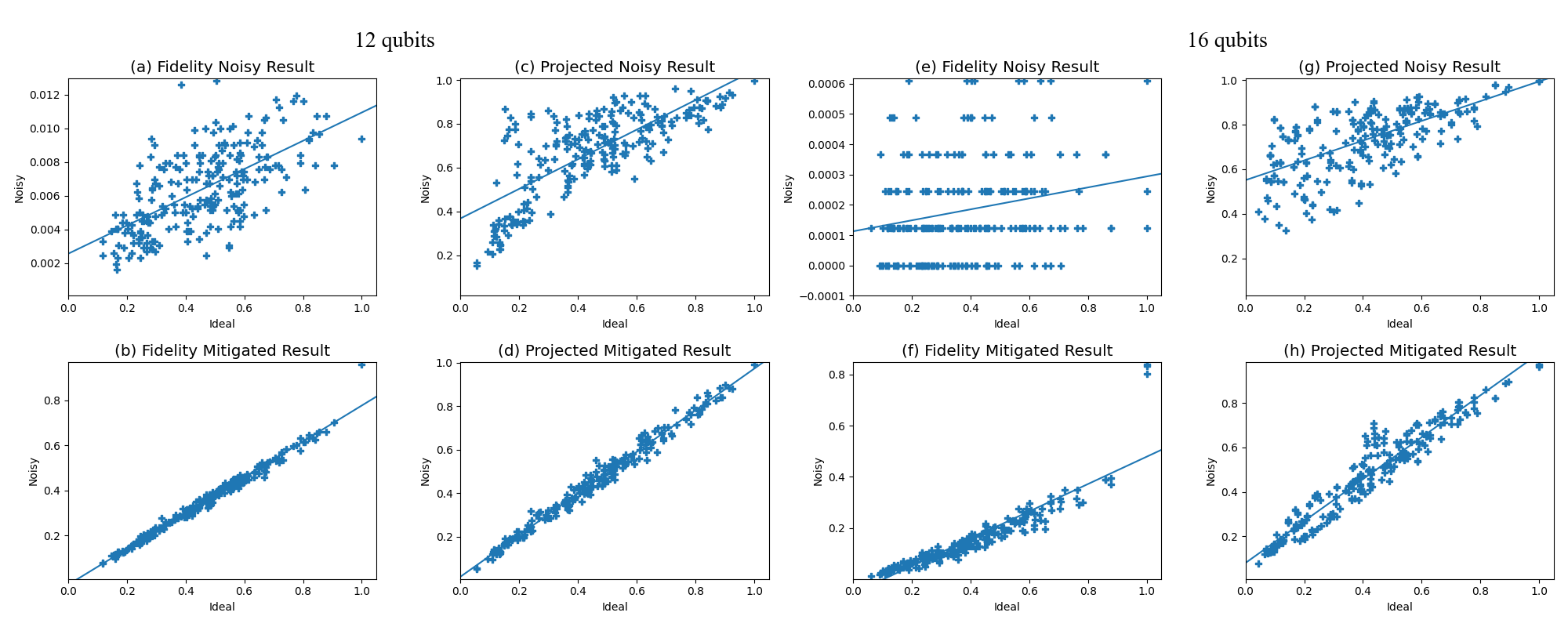}
    \caption{\label{Fig:qke_hardware_result_qubits12-16} 
    Hardware results of fidelity and projected quantum kernels are shown. The results are obtained using the Z-ZZ-feature map with reps=1, $\alpha$=2, entanglement='linear' and data map function=$\phi(x,y)=(\pi-x)(\pi-y)$.
    The results of 12 qubits and 16 qubits are show in figures (a$\sim$d) and (e$\sim$h), respectively. The ideal simulation is plotted on the vertical axis, and the result of the hardware is on horizontal axis.
    }
\end{figure*}

In this section we compare the performance of SVM models built using quantum kernels computed on an IBM quantum computer, $ibm\_auckland$. The Bank Marketing dataset is used in this section.
The performance of the error mitigation technique proposed in this paper, measured at 12 and 16 qubits, is plotted in Fig.~\ref{Fig:qke_hardware_result_qubits12-16}. This figure compares the hardware runs (horizonal axis) with ideal simulations for both fidelity and projected quantum kernels. Comparison with the ideal kernel is possible as the number of qubits remains in the simulatable regime. 11 data points giving kernel magnitudes that are evenly distributed in the range between 0 and 1 are chosen. The dataset is standardized following the procedure described in the simulator section. We used 8192 shots, and 16 random Pauli operators for Pauli twirling. 2000 shots per qubit were used for measurement error calibration.

In our analysis we set result of the simulation and the hardware as explanatory and response variable, respectively, and we perform linear regression. A perfect fit will get the slope of 1 and $r^2=1$. 
The error mitigated (EM) and unmitigated (noisy) results are plotted in the bottom and the top row, respectively. Focusing on the top, we see that the performance of both fidelity and projected kernels declines with increasing the number of qubits. The $r^2$ of fidelity kernels drops from 0.410 to 0.048 in the investigated range. Similarly the $r^2$ of projected kernels drops from 0.567 to 0.432. Therefore, the drop in performance is more pronounced for the fidelity case. In fact, the magnitudes of fidelity kernel elements are concentrating towards a small number in both 12 and 16 qubit case (Fig.~\ref{Fig:qke_hardware_result_qubits12-16}e). 
This relates, once again, to the exponential concentration of kernel with respect to feature size discussed in \cite{thanasilp2022exponential} and Sec.\ref{exponential_concentration}.
We see a significant boost in performance with error mitigation (bottom row of Fig.~\ref{Fig:qke_hardware_result_qubits12-16}). The $r^2$ of fidelity kernels for 12 and 16 qubit calculations are 0.984 and 0.829, respectively, and the mean of kernel magnitudes are relatively large. The $r^2$ of projected kernels for 12 and 16 qubits are 0.984 and 0.926, respectively, and kernel values are more consistent with ideal values. Therefore, our result favors the projected kernel approach.
In addition, this demonstrates the effectiveness of the error mitigation and suppression pipeline presented in this report.

Table.~\ref{tb:fidelity_projected_hardware_results} gives a further comparison of the fidelity and projected quantum kernel. 
It confirms the effectiveness of our error mitigation pipeline for both types of kernel, giving the $r^2$ value above 0.8 in the two cases at the largest qubit size of 20.
In contrast, we are unable to obtain a meaningful result beyond 4 qubits without error mitigation for this quantum kernel. At 8 qubits $r^2$ of the noisy result is around $0.6$ for both kernels. Furthermore, the slope of the fidelity kernel is large ($>9$) for qubit size greater than $4$, suggesting, once again, a concentration of kernel values without error mitigation for this kernel.

\begin{table}[htbp]
    \centering
    \begin{tabular}{|c|c|c|c|c|} \hline
        Kernel type & qubits & Error migiation & Slope & $r^2$ \\ \hline
        \multirow{10}{*}{Fidelity} & \multirow{2}{*}{4} & noisy & $1.236$ & $0.952$   \\ \cline{3-5}
        & & EM & $0.9716$ & $0.997$ \\ \cline{2-5}
        & \multirow{2}{*}{8} & noisy & $9.145$ & $0.671$ \\ \cline{3-5}
        & & EM & $1.024$ & $0.995$ \\ \cline{2-5}
        & \multirow{2}{*}{12} & noisy & $48.86$ & $0.410$ \\ \cline{3-5}
        & & EM & $1.237$ & $0.984$ \\ \cline{2-5}
        & \multirow{2}{*}{16} & noisy & $264$ & $0.048$ \\ \cline{3-5}
        & & EM & $1.554$ & $0.829$ \\ \cline{2-5}
        & \multirow{2}{*}{20} & noisy & $300.3$ & $0.013$ \\ \cline{3-5}
        & & EM & $5.207$ & $0.905$ \\ \cline{2-5} \hline
        \multirow{10}{*}{Projected} & \multirow{2}{*}{4} & noisy & $1.075$ & $0.965$  \\ \cline{3-5}
        & & EM & $0.9888$ & $0.999$ \\ \cline{2-5}
        & \multirow{2}{*}{8} & noisy & $1.187$ & $0.563 $ \\ \cline{3-5}
        & & EM & $1.007$ & $0.997$ \\ \cline{2-5}
        & \multirow{2}{*}{12} & noisy & $0.8397$ & $0.567$ \\ \cline{3-5}
        & & EM & $1.029$ & $0.984$ \\ \cline{2-5}
        & \multirow{2}{*}{16} & noisy & $0.974$ & $0.432$ \\ \cline{3-5}
        & & EM & $0.9808$ & $0.926$ \\ \cline{2-5}
        & \multirow{2}{*}{20} & noisy & $0.8582$ & $0.648$ \\ \cline{3-5}
        & & EM & $0.8447$ & $0.812$ \\ \cline{2-5} \hline
    \end{tabular}
    \caption{
    We tabulate the linear regression metrics used to compare the quality of fidelity and projected quantum kernels with and without error mitigation. Here the result of the simulation and the hardware are used as explanatory and response variable, respectively, and linear regression fit is performed to measure quality. The kernel parameters used in  Fig.~\ref{Fig:qke_hardware_result_qubits12-16} is employed.
    }
    \label{tb:fidelity_projected_hardware_results}
\end{table}

We now compare the performance of SVM models built using the quantum kernels computed on the hardware. Here the total dataset size is $100$, and $30$ is used as a test data. The combination of $P_i$ (in Eq.~\ref{eq:feature_map_diag_gate}) and $\alpha$ included in the MKL models are shown in Table.~\ref{table:mkl_linear_fit}. Considering the resource scaling of the fidelity and the projected quantum kernels, which are $O(n^2)$ and $O(3nm)$, respectively, we limit our investigation to the projected kernels going forward. Note that $n$ is the number of data points and $m$ is the number of features. In addition, the performance of projected kernels are more consistent with ideal values as demonstrated earlier in this section.

First, we tabulate the linear regression metrics in table.~\ref{table:mkl_linear_fit}, where the slope and $r^2$ are computed for all combinations of $P_i$ and $\alpha$ used to build the MKL model. 
The results obtained with and without error mitigation are shown for comparison. We see a performance improvement with error mitigation in most cases.

\begin{table*}
    \centering
    \begin{tabular}{|l|l|l|ll|ll|ll|}
        \hline
        ~ & ~ & ~ & \multicolumn{2}{l|}{8 qubits} & \multicolumn{2}{l|}{12 qubits} & \multicolumn{2}{l|}{16 qubits} \\
        \cline{4-9}
        $P_i$ & $\alpha$ & Error Mitigation & Slope & r$^2$ & Slope & r$^2$ & Slope & r$^2$ \\ \hline
        \multirow{4}{*}{Y} & \multirow{2}{*}{0.2} & noisy & 1.097 & 0.981 & 1.017 & 0.990 & 1.056 & 0.986 \\ \cline{3-9}
        ~ & ~ & EM & 0.924 & 0.987 & 0.583 & 0.863 & 0.957 & 0.973 \\ \cline{2-9}
        ~ & \multirow{2}{*}{0.3} & noisy & 1.068 & 0.995 & 1.052 & 0.995 & 1.039 & 0.994 \\ \cline{3-9}
        ~ & ~ & EM & 0.893 & 0.995 & 0.778 & 0.986 & 0.860 & 0.993 \\ \cline{1-9}
        \multirow{4}{*}{ZZ} & \multirow{2}{*}{0.7} & noisy & 1.876 & 0.699 & 0.941 & 0.695 & 1.569 & 0.923 \\ \cline{3-9}
        ~ & ~ & EM & 1.018 & 0.965 & 1.060 & 0.973 & 1.099 & 0.863 \\ \cline{2-9}
        ~ & \multirow{2}{*}{0.8} & noisy & 0.834 & 0.675 & 1.119 & 0.753 & 1.643 & 0.910 \\ \cline{3-9}
        ~ & ~ & EM & 1.018 & 0.993 & 1.055 & 0.978 & 1.125 & 0.919 \\ \cline{1-9}
        \multirow{4}{*}{X-ZZ} & \multirow{2}{*}{0.7} & noisy & 0.928 & 0.653 & 1.245 & 0.695 & 1.954 & 0.915 \\ \cline{3-9}
        ~ & ~ & EM & 0.827 & 0.967 & 0.964 & 0.964 & 1.060 & 0.856 \\ \cline{2-9}
        ~ & \multirow{2}{*}{0.8} & noisy & 1.269 & 0.745 & 1.036 & 0.760 & 1.576 & 0.899 \\ \cline{3-9}
        ~ & ~ & EM & 1.047 & 0.978 & 1.043 & 0.975 & 1.090 & 0.942 \\ \cline{1-9}
        \multirow{2}{*}{Y-ZZ} & \multirow{2}{*}{0.4} & noisy & 1.381 & 0.966 & 1.197 & 0.964 & 1.347 & 0.949 \\ \cline{3-9}
        ~ & ~ & EM & 1.097 & 0.992 & 1.058 & 0.943 & 1.213 & 0.876 \\ \cline{1-9}
        \multirow{2}{*}{Z-ZZ} & \multirow{2}{*}{0.4} & noisy & 1.196 & 0.920 & 1.446 & 0.867 & 1.604 & 0.623 \\ \cline{3-9}
        ~ & ~ & EM & 1.096 & 0.964 & 0.986 & 0.888 & 0.960 & 0.700 \\ \hline
    \end{tabular}
    \caption{
    We tabulate the linear regression metrics with (EM) and without (noisy) error mitigation. Results are tabulated for combinations of $P_i$ (in Eq.~\ref{eq:feature_map_diag_gate}) and $\alpha$ used to build an MKL model in this report.
    }
    \label{table:mkl_linear_fit}
\end{table*}

We find there is no single learner approach that performs well for all qubit sizes.
\textit{MKL models on hardware are more robust}.In table.~\ref{table:mkl_hw_auc} we tabulate the performance of single and multiple kernel learning approaches considered in this report, and we demonstrate the benefit of MKL when implemented on a hardware.
We see that the performance of single learners are sporadic on hardware. See for example the case for $P_i=$Y-ZZ and $\alpha=0.4$, where the model performance for both train and test set is good for the 8 and 12 qubit case. The results are also consistent with the ideal simulation. However, the ROC-AUC of the model drops to 0.525 and 0.716 at 16 qubits for the train and test set, respectively.  ROC-AUC for the train and test sets are consistent and above $0.90$ in most cases. The results are also more consistent with the ideal calculation. We note an exception for MKL SDP at 16 qubits, where the train and test ROC-AUC of the hardware are 0.750 and 0.897, respectively.
The result leads us to emphasize the benefit of MKL when implemented on a hardware.

\begin{table*}
    \centering
    \begin{tabular}{|l|l|l|l|l|l|l|l|l|l|l|l|l|}
    \hline
        ~ & \multicolumn{4}{l|}{8 qubits} & \multicolumn{4}{l|}{12 qubits} & \multicolumn{4}{l|}{16 qubits} \\ \hline
        ~ & \multicolumn{2}{l|}{Ideal} & \multicolumn{2}{l|}{EM} & \multicolumn{2}{l|}{Ideal} & \multicolumn{2}{l|}{EM} & \multicolumn{2}{l|}{Ideal} & \multicolumn{2}{l|}{EM} \\ \hline
        kernel & Train & Test & Train & Test & Train & Test & Train & Test & Train & Test & Train & Test \\ \hline
        ('Y', 0.2) & 0.954 & 0.741 & 0.869 & 0.422 & 0.751 & 0.491 & 0.758 & 0.776 & 0.719 & 0.388 & 0.500 & 0.750 \\ \hline
        ('Y', 0.3) & 0.977 & 0.922 & 0.957 & 0.802 & 0.750 & 0.966 & 0.750 & 0.966 & 0.721 & 0.569 & 0.750 & 0.897 \\ \hline
        ('ZZ', 0.8) & 0.998 & 0.966 & 0.991 & 0.991 & 0.918 & 0.983 & 0.873 & 0.948 & 0.991 & 0.974 & 0.922 & 0.707 \\ \hline
        ('ZZ', 0.7) & 1.000 & 0.991 & 0.989 & 0.905 & 0.909 & 0.983 & 0.999 & 0.767 & 0.914 & 0.888 & 0.969 & 0.190 \\ \hline
        ('X-ZZ', 0.7) & 1.000 & 0.991 & 1.000 & 0.940 & 0.909 & 0.983 & 0.981 & 0.862 & 0.914 & 0.888 & 1.000 & 0.603 \\ \hline
        ('X-ZZ', 0.8) & 0.998 & 0.966 & 0.885 & 0.914 & 0.918 & 0.983 & 0.928 & 0.983 & 0.991 & 0.974 & 0.712 & 0.983 \\ \hline
        ('Y-ZZ', 0.4) & 0.980 & 0.983 & 0.975 & 0.966 & 0.999 & 0.966 & 1.000 & 0.957 & 1.000 & 0.991 & 0.525 & 0.716 \\ \hline
        ('Z-ZZ', 0.4) & 0.981 & 0.974 & 0.980 & 0.948 & 0.997 & 0.957 & 0.750 & 0.974 & 0.721 & 0.836 & 1.000 & 0.517 \\ \hline
        MKL AVE & 0.997 & 0.991 & 0.998 & 0.991 & 0.999 & 0.966 & 1.000 & 0.940 & 1.000 & 1.000 & 1.000 & 0.931 \\ \hline
        MKL PROJ & 0.980 & 0.983 & 0.983 & 0.974 & 0.998 & 0.957 & 1.000 & 0.974 & 1.000 & 0.991 & 1.000 & 0.983 \\ \hline
        MKL SDP & 0.980 & 0.983 & 0.975 & 0.966 & 0.999 & 0.966 & 1.000 & 0.966 & 1.000 & 0.991 & 0.750 & 0.897 \\ \hline
        MKL CENT & 0.960 & 0.905 & 0.994 & 1.000 & 1.000 & 0.974 & 1.000 & 0.845 & 0.955 & 1.000 & 1.000 & 0.931 \\ \hline
    \end{tabular}
    \caption{
    We tabulate the model performance, measured in ROC-AUC, of SVM models built using various single learners and MKL kernels. The kernels are obtained using both simulator (Ideal) and hardware with the error mitigation pipeline (EM). Results are tabulated for combinations of $P_i$ (in Eq.~\ref{eq:feature_map_diag_gate}) and $\alpha$ and MKL models built from all combinations. The target-kernel alignment schemes discussed in Sec.~\ref{kernel_alignment}, AVE, PROJ, SDP and CENT, are employed to optimize the weights of MKL.
    }
    \label{table:mkl_hw_auc}
\end{table*}

\section{Conclusions}

Quantum support vector machine is a promising near term candidate to offer uplift in binary classification tasks, however, picking the appropriate kernel to describe the data remains challenging, particularly when data structure is largely unknown. Further instabilities arise from exponentially concentrated kernels due to scale and noise that inhibit sufficient training. We proposed an approach using a linear combination of kernels whose weights are determined by a classical optimizer. The optimizer routine aligns the combined kernel with the target kernel for the training set. This approach doesn't suffer from issues stemming from parameterized quantum circuits in QKA. The approach is evaluated on data sets relevant to the financial services industry.

QMKL is tested both in simulation and on IBM's quantum hardware. The method showed advantage in ROC-AUC scores for two out of the three data sets, and QML produced higher quality discrimination scores for a majority of samples. Quite interestingly, QMKL demonstrated advantage on the HSBC Digital Payment data showing promise for industrial application and integration of QML routines in fraud detection workflows. We find the QMKL method stabilizes kernel variance and mean making it robust against exponential concentration with larger qubit dimensions. Linear regression metrics were used to compare the quality of the fidelity quantum kernel approach and the projected quantum kernel approach on hardware both with and without error mitigation. The projected quantum kernel yielded more consistent slopes and $r^2$ scores, therefore, the projected method was used for the QML algorithm. We compared SVM performance between single quantum kernels and multiple quantum kernels. The results show that QMKL consistently performs better as more qubits are added with up to an average 12.5\% improvement compared to the single quantum kernel for the 16 qubit case.

Our results demonstrate the impact of QMKL through performance gain not just in simulation, but on quantum hardware. Further improvements to QMKL are needed to provide truly meaningful use in classification with the addition of a larger feature space (e.g. qubits) and larger training data sets (e.g. parallel processing of kernel elements in a quantum super computing center). Performance gains are vary by data set, however, we do find significant results to substantiate further application research of scaled quantum machine learning algorithms.
 
\section*{Acknowledgements}
DISLCAIMER: This paper was prepared for information purposes, and is not a product of HSBC or its affiliates. Neither HSBC nor any of its affiliates make any explicit or implied representation or warranty and none of them accept any liability in connection with this paper, including, but not limited to, the completeness, accuracy, reliability of information contained herein and the potential legal, compliance, tax or accounting effects thereof. Copyright HSBC Group 2023.

\bibliographystyle{ieeetr}
\bibliography{references}

\end{document}